\documentclass[twocolumn,showpacs,preprintnumbers,amsmath,amssymb,superscriptaddress,10pt,aps,prb]{revtex4-2}
\usepackage{graphicx}
\usepackage{dcolumn}% Align table columns on decimal point
\usepackage{bm}% bold math
\usepackage{amsmath}
\usepackage{times}
\usepackage{color}
\usepackage[breaklinks=true,colorlinks,citecolor=blue,linkcolor=blue,urlcolor=blue]{hyperref}

\makeatletter

\newcommand{\Rmnum}[1]{\expandafter\@slowromancap\romannumeral #1@}
\makeatother

\begin{document}

\title{Observation of Fermi liquid phase with broken symmetry in a single crystalline nanorod of Pr$_2$Ir$_2$O$_7$}
\author{Bikash Ghosh}
\author{Abhishek Juyal}
\author{Sourav Biswas}
\affiliation{Department of Physics, Indian Institute of Technology Kanpur, Kanpur 208016, India}
\author{R. Rawat}
\affiliation{UGC-DAE Consortium for Scientific Research, University Campus, Khandwa Road, Indore 452001, India}
\author{Arijit Kundu}
\affiliation{Department of Physics, Indian Institute of Technology Kanpur, Kanpur 208016, India}
\author{Soumik Mukhopadhyay}
\email{soumikm@iitk.ac.in}
\affiliation{Department of Physics, Indian Institute of Technology Kanpur, Kanpur 208016, India}

\begin{abstract}
We report experimental evidence of emergent broken symmetry Fermi liquid state in an isolated single crystalline nanorod of $\rm Pr_2 Ir_2 O_7$. We find clear signature of the onset of the Fermi liquid behavior at low temperature marked by the sign inversion of magnetoresistance from negative at high temperature, characteristic of incoherent Kondo scattering, to positive as well as a $\rm T^2$ dependence of resistivity at low temperature. A resistive anomaly is observed, which is accompanied by thermal hysteresis in the presence of magnetic field, suggesting itinerant metamagnetism. The observed high field negative magnetoresistance with quadratic field dependence at low temperature, which is most likely due to suppression of itinerant spin fluctuation, and the irreversibility of the magneto-resistive properties in the Fermi liquid regime suggest existence of an unusual state with broken spin rotation and time reversal symmetry, hallmark of `hastatic' order. The major features of such temperature dependence of resistivity and magnetoresistance can be explained in a phenomenological model incorporating two distinct hybridization channels, which is physically consistent with the possibility of the formation of the `hastatic' Fermi liquid phase.
\end{abstract}

%\pacs{75.47.Lx, 72.20.-i, 79.60.-i, 71.70.Ej, 78.70.Dm}

\maketitle

For the non-Kramers ion with an even number of electrons, the two channel Kondo effect leads to spinorial hybridization, resulting potentially in a phase transition to a new symmetry breaking state called `hastatic' order~\cite{chandra}. The idea of hastatic order was originally introduced for $\rm URu_2 Si_2$ to explain the observed `hidden' order  phase~\cite{chandra}, provisionally `hidden' since the order parameter remains experimentally undetected till date. However, the $5$f electron in U ion is partially localized and partially itinerant. The crystalline electric field levels are often too broad to be identified. On the contrary, the $4$f electron in non-Kramers Pr ion is highly localized and has well defined crystal field levels, making it a more suitable candidate to exhibit two channel spinorial hybridization. Despite the theoretical plausibility, the hastatic order in Pr based systems still remains largely unexplored. A family of PrT$_{2}$X$_{20}$ (T=transition metal, X= Al, Zn and Cd) compounds with non-Kramers doublet ground state provides key prerequisites for the two channel Kondo effect and has recently drawn considerable attention~\cite{Tsujimoto,Onimaru,Onimaru1,Sakai,Onimaru2,Yoshida}. Specifically, in $\rm Pr Ir_{2} Zn_{20}$ with Fd$\bar{3}$m symmetry, the quadrupolar order is suppressed by the magnetic field B$_{c}$ and in the vicinity of B$_{c}$, pronounced anomalies in specific heat~\cite{Onimaru}, Seebeck coefficient~\cite{Ikeura}, elastic constant and peculiar Fermi liquid state are observed~\cite{Onimaru2}. PrPb$_{3}$ is another non-kramers material that shows some promise of the heavy fermion behavior within the quadrupolar ordered state at high field, making it a hastatic order candidate~\cite{Onimaru3, Morin}.

Among the Pyrochlore Iridates~\cite{Wan, Nakatsuji, Kim, Pesin, Yang, Kargarian, Machida, William, Go, Cheng}, $\rm Pr_2 Ir_2 O_7$ is supposed to be a metallic `chiral spin liquid' with macroscopically broken time reversal symmetry but no magnetic long range order down to the lowest temperature~\cite{Machida, Tokiwa, Nakatsuji}. The low temperature transport and magnetic properties are attributed to the `inter-site' Kondo coupling mechanism~\cite{Nakatsuji3}, similar to a two-impurity Kondo problem described by a low energy effective Hamiltonian where the RKKY interaction between $\rm Pr^{3+}$ local moments mediated by $\rm Ir^{4+}$ conduction electrons dominates over the effective Kondo coupling between the local moments and the conduction electrons. However, such a straightforward application of the physics of Kondo lattice may not necessarily give the complete picture~\cite{Udagawa}. The local $\rm Pr^{3+} 4f^2$ moments have non-Kramers $J\rm=4$ doublet ground state. Recently Rau et. al. proposed~\cite{Rau} that the hybridization between the local Pr pseudospins and Ir conduction electrons could break spatial as well as time reversal symmetries in $\rm Pr_2 Ir_2 O_7$, similar to the `hidden order' transition in $\rm URu_2 Si_2$~\cite{Mydosh, Flint}. Such symmetry breaking phase transition is also associated with heavy Fermi liquid formation~\cite{Senthil} . However, bulk single crystalline $ \rm Pr_2 Ir_2 O_7 $ does not show any evidence of such behavior~\cite{Nakatsuji}.

Recent discovery of quadratic Fermi node in the angle resolved photoemission spectroscopy (ARPES) study introduced new dimensions to the debate as the system is now thought to be unstable towards forming a large variety of strongly correlated quantum phases such as topological Mott insulator, magnetic Weyl semimetal (WSM), quantum spin and anomalous Hall states due to the strain and confinement effects~\cite{Kondo}. Such novel `Non-Fermi Liquid' (NFL) phases with high dielectric constant and appearance of Drude peak at low temperature on one hand, and a magnetic WSM phase on the other, have recently been reported in $ \rm Pr_2 Ir_2 O_7 $ thin film with tensile strain along the surface normal [111] direction~\cite{Cheng, Ohtsuki}.

It is thus instructive to study the properties of $ \rm Pr_2 Ir_2 O_7 $ in different sample geometries. The study of electrical properties of $ \rm Pr_2 Ir_2 O_7 $ in the nanorod geometry can potentially lead to important insights on the interplay of Kondo coupling, RKKY interaction and frustration parameter. A small change in lattice parameter without altering the space group symmetry or distortion in the lattice geometry can change the balance between the energy scales and the frustration parameter, thereby providing a natural control parameter for non-thermal tuning of the quantum phases~\cite{Tokiwa, Kondo}. In this paper we discuss the electrical transport properties of an isolated single crystalline nanorod based on Pr$_2$Ir$_2$O$_7$ and the emergence of an unusual Fermi liquid state associated with a symmetry breaking phase transition.
\begin{figure}
\includegraphics[width=\linewidth]{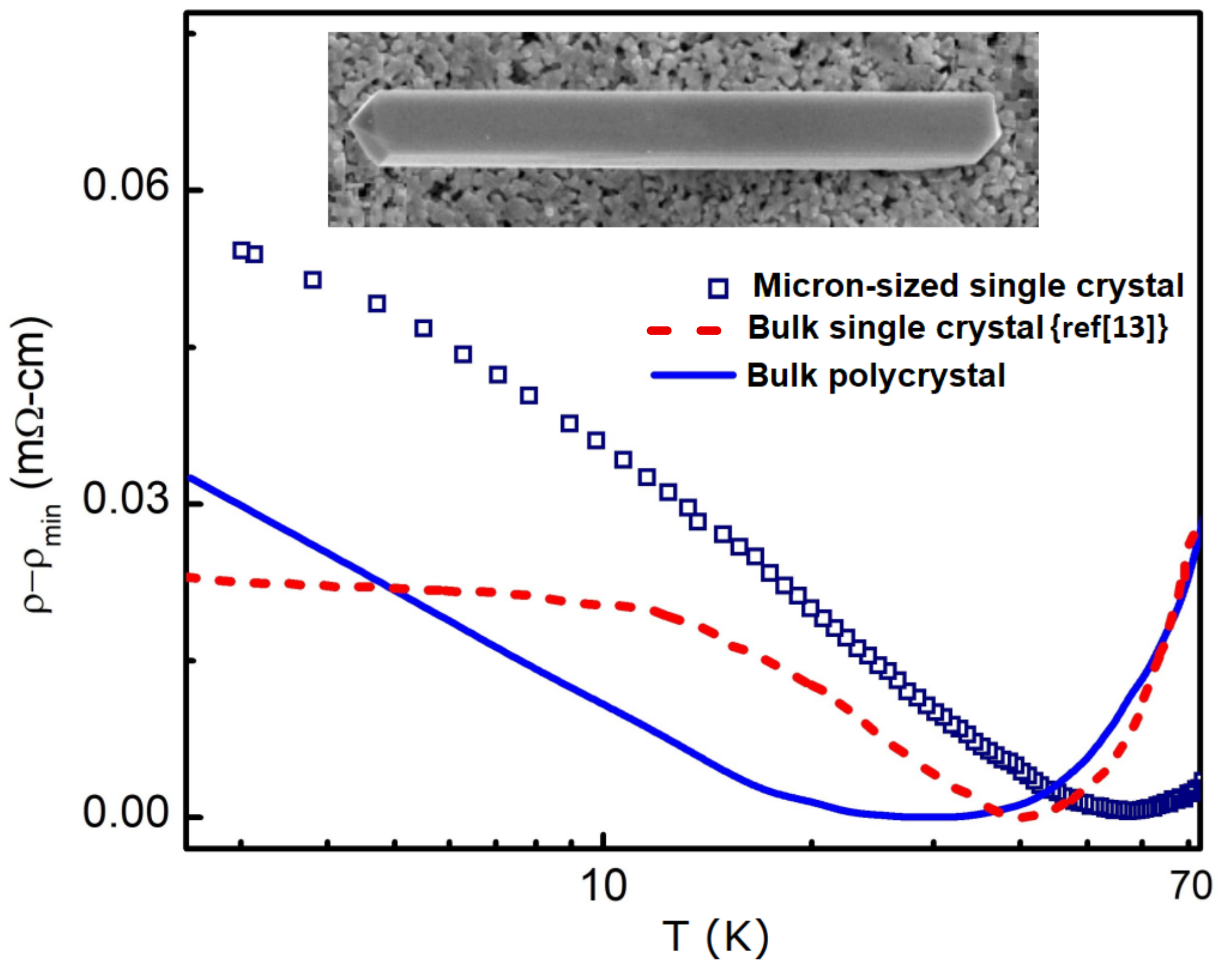}
\caption{The temperature dependence of spin contribution to resistivity of the bulk polycrystalline sample and PIO$_{\rm ref}$ (single crystalline rod with $15$ micron width and $100$ micron length) are compared with that of bulk single crystalline sample reported in literature. Inset: SEM image of the PIO$_{\rm ref}$ sample.}\label{fig1}
\end{figure}

Starting with the bulk polycrystalline sample based on Pr$_2$Ir$_2$O$_7$ (PIO) prepared using solid-state reaction method~\cite{Dwivedi, Bikash1}, single crystalline nanorods with cross section varying between $0.01-0.04\,\mu$m$^2$ and micro-crystal rods with cross-section varying between $50-1000\, \mu$m$^2$ (henceforth to be treated as a reference sample and called PIO$_{\rm ref}$) were prepared using self-flux method. Details of sample preparation and characterization are discussed in the Appendix A. XPS measurements could be performed on the PIO$_{\rm ref}$ samples. As compared to the parent bulk polycrystalline sample, no significant change in Ir local environment was observed in PIO$_{\rm ref}$. However, the local environment around $\rm Pr^{3+}$ undergoes marginal modification (see Appendix).

We shall discuss the electrical transport properties of the nanorod later. For the time being, let us focus on the PIO$_{\rm ref}$ samples. Electrical resistivity measurements were performed on a number of such micro-crystal rods. In Fig.~\ref{fig1}, we compare the resistivity upturn in a representative PIO$_{\rm ref}$ sample to the same for the bulk polycrystalline parent sample and the `bulk' single crystalline sample reported before~\cite{Nakatsuji}. Although the resistivity minimum of PIO$_{\rm ref}$ is shifted towards higher temperature at $\rm T=58$ K compared to the bulk single crystal, the resistivity upturn begins to saturate at a lower temperature. As a result, the Kondo temperature $\rm T_K$ extracted from fitting of the experimental data with Hamann's expression~\cite{Hamann} turns out to be $20$ K, close to the $\rm T_K$ reported for bulk single crystal. The resistivity upturn or presumably the Kondo contribution to resistivity is higher in PIO$_{\rm ref}$ compared to the bulk single crystal or the poly-crystalline sample (Fig.~\ref{fig1}). The residual resistivity of the bulk polycrystalline sample and PIO$_{\rm ref}$ were estimated using Hamann's expression. The residual resistivity of the former turns out to be $0.92\, \rm m\Omega-$cm while the same for PIO$_{\rm ref}$ has similar value to the one reported by Nakatsuji \textit{et al}.~\cite{Nakatsuji}.

\begin{figure}
\includegraphics[width=\linewidth]{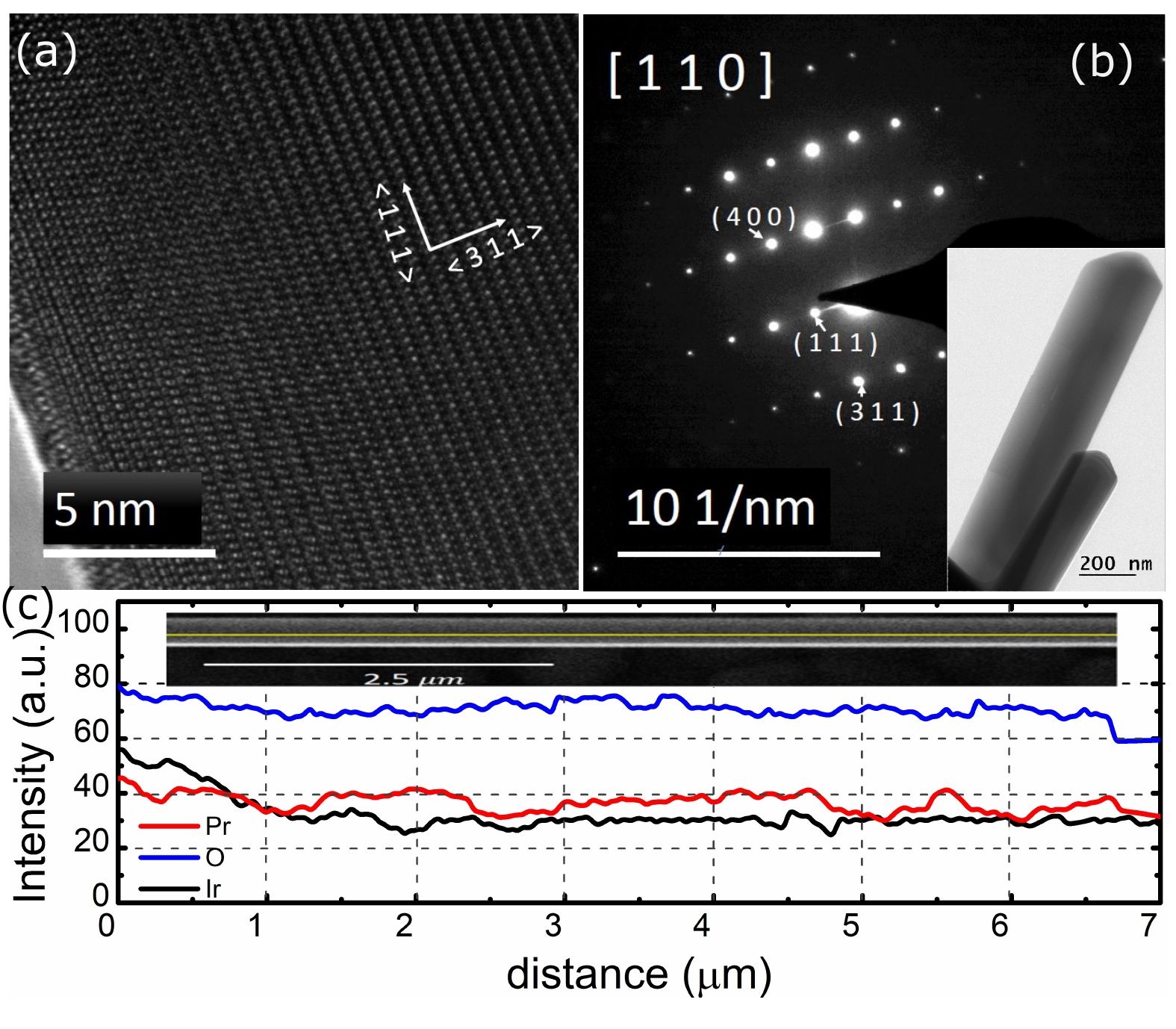}
\caption{(a) The real space HRTEM image and (b) corresponding SAED pattern for the nanorod is shown. (c) The homogeneity of the chemical composition is shown across the length of the nanorod using EDS.}\label{fig2}
\end{figure}

Single-crystalline nanorods of PIO were drop cast on a silicon oxide substrate with subsequent metallization using e-beam lithography~\cite{Abhishek1,Abhishek2}. For details of device fabrication and characterization, see the Appendix. High resolution transmission electron microscopy (HRTEM) in real space and selected area electron diffraction (SAED) shows that the nanorods are single crystalline with unchanged space group symmetry compared to the bulk and with growth direction along [111] (discussed in the Appendix). A number of energy dispersive spectroscopy (EDS) line scan profile for a PIO nanorod gives an average atomic percentage ratio ${\rm Pr}/{\rm Ir} \sim 0.94$ (Fig.~\ref{fig2}c). We also find from HRTEM and SAED pattern (Fig.~\ref{fig2}a, b) that the lattice parameter in the nanorod is $10.22 \rm \AA$, reduced by $1.6\%$ compared to the bulk system.

\begin{figure}
\includegraphics[width=\linewidth]{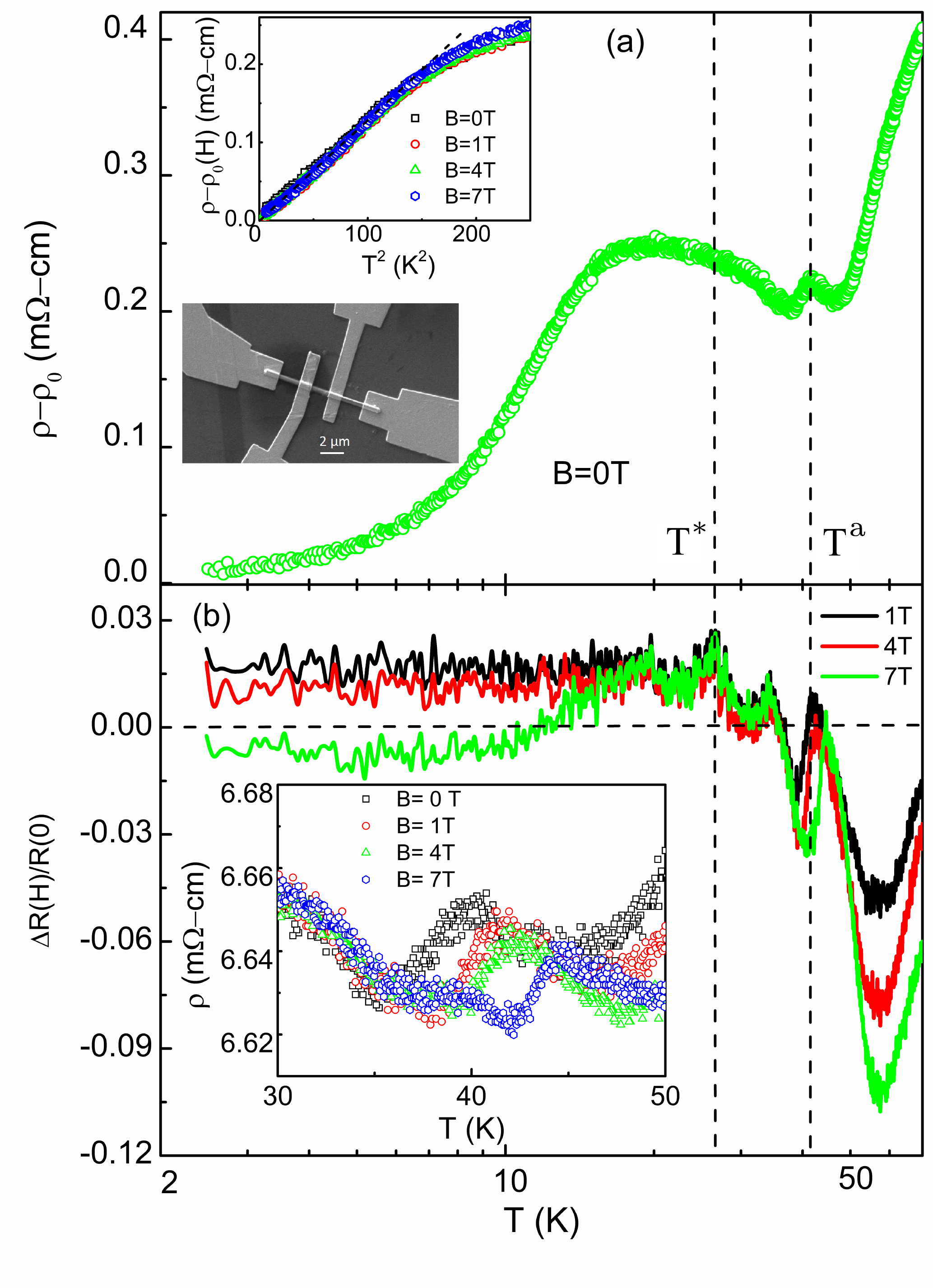}
\caption{(a) The temperature dependence of resistivity of the nanorod in absence of magnetic field. Upper inset: At low temperature, $T^2$ dependence of resistivity is observed, which is unaffected by application of magnetic field. Lower inset is an image of the nanorod device for electrical measurements. (b) The corresponding temperature dependence of MR at different magnetic fields. Inset: The shift of the resistive anomaly with magnetic field.}\label{fig3}
\end{figure}

The temperature dependence of resistivity of the nanorod of diameter $150$ nm, measured down to $2$~K (Fig.~\ref{fig3}a) has several interesting features: $1)$ There is a clear resistive anomaly at $\rm T^{\rm a}=42$~K. The `$\Lambda$' shaped anomaly  progressively shifts towards higher temperature on application of magnetic field without getting suppressed (Inset, Fig.~\ref{fig3}b). $2)$ On further lowering of temperature we observe a shallow resistivity upturn which eventually flattens off near $\rm T^\ast=26$~K. $3)$ On the low temperature side we observe metallic temperature dependence of resistivity. Below $10$~K, the temperature dependence is quadratic, with the exponent largely remaining non-responsive to magnetic field (Inset, Fig.~\ref{fig3}a). We have also extracted the exponent from the log-log plot of $\rho - \rho_0$ vs. $\rm T$ at different magnetic field ($\mathrm{B}$), which turns out to be $1.80\pm 0.02$ (not shown in the figure). We can safely conclude that this is characteristic of a coherent Fermi liquid state. The residual resistivity of the single crystal nanorod is $6.4\, \rm m\Omega-$cm, slightly higher than that reported for thin film samples of thickness $100$ nm with c axis along [111] direction~\cite{Cheng}.

\begin{figure}
\includegraphics[width=\linewidth]{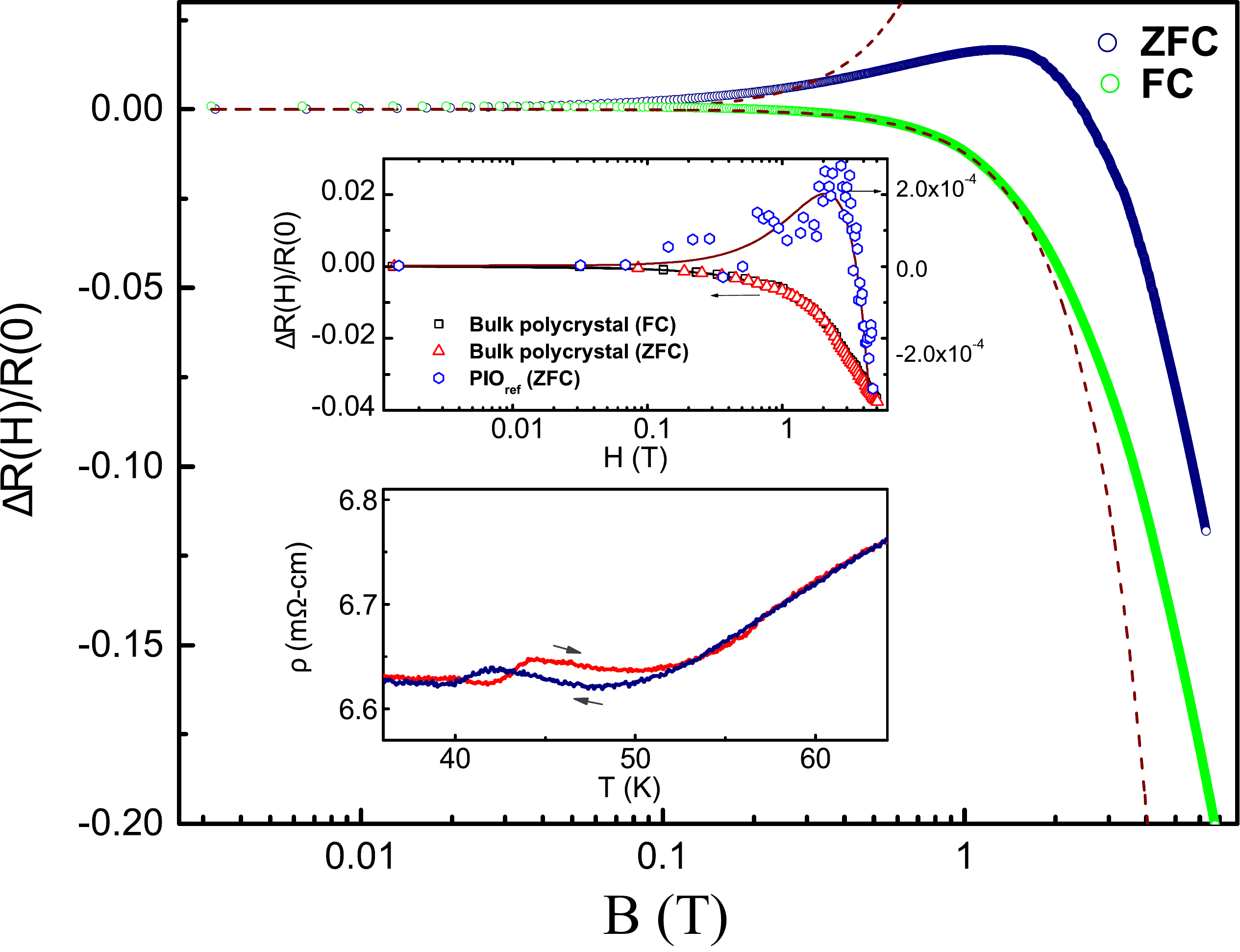}
\caption{Magnetic field dependence of MR for the nanorod at $1.5$ K showing irreversibility of MR under FC and ZFC protocol. The dashed line shows $\rm H^2$ variation of MR. Upper inset: PIO$_{\rm ref}$ shows positive MR at low magnetic field followed by negative MR at higher magnetic field. The MR for bulk polycrystalline sample is negative and does not show any irreversibility in the ZFC and FC curve. Lower inset: Thermal hysteresis around the resistive anomaly for the nanorod at $4$~T.}\label{fig4}
\end{figure}

In Fig.~\ref{fig3}b, we plot the temperature dependence of magnetoresistance (MR) at different constant magnetic fields. The MR turns out to be strikingly different in the two temperature regimes separated by $\rm T^\ast$. While below $\rm T^\ast$, a small positive MR is observed, which remains independent of temperature, MR above $\rm T^\ast$ turns out to be negative with the following features: $1)$ A sharp extremum in the negative MR is observed around $\rm T^{\rm a}$ which shifts towards higher temperature with application of strong magnetic field. $2)$ Above $\rm T^{\rm a}$, a large negative MR is observed which increases with application of magnetic field, characteristic of inelastic Kondo spin flip scattering. As the temperature drops below $\rm T\sim \rm T^\ast$, we observe a resistivity which is quadratic in $\rm T$ and $\rm H$ (as shown by the dashed line in ZFC MR in Fig.~\ref{fig4}). Such properties of the nanorod are in striking contrast to that observed in the bulk single crystalline sample~\cite{Nakatsuji} or even in micron sized rods such as PIO$_{\rm ref}$ for the present study.

Curiously, the MR at high magnetic field becomes negative below $10$ K, roughly the regime where the $\rm T^2$ dependence is followed (Fig.~\ref{fig3}b). The field dependence of MR at low temperature has another interesting property: irreversibility in ZFC and FC MR. While the ZFC MR is positive and quadratic in $\rm H$ at low field, and negative at higher field, the FC MR remains negative throughout and quadratic in $\rm H$ (Fig.~\ref{fig4}). Such irreversibility between FC and ZFC MR could be due to the existence of an unusual Fermi liquid state with broken time reversal symmetry similar to URu$_2$Si$_2$~\cite{urs}. The field dependence of MR for the polycrystalline sample and the PIO$_{\rm ref}$ is shown in one of the insets of Fig.~\ref{fig4}. While polycrystalline sample does not show any irreversibility in MR and remains negative, the MR for PIO$_{\rm ref}$ is positive at low field and negative at higher field.

The $\Lambda$ shaped anomaly and its sensitivity to magnetic field is similar in nature to the ones observed in heavy Fermion Kondo systems such as CeRu$_2$Si$_2$~\cite{Kambe}. In addition to the shift of the resistive anomaly with magnetic field, we observe a magnetic field induced thermal hysteresis of resistivity near the anomaly at temperature $\rm T^{\rm a}$ (lower inset, Fig.~\ref{fig4}). Similar hysteresis has also been observed in $\rm U Ru_2 Si_2$~\cite{Kim1}. This is attributed to itinerant metamagnetism, the magnetic field induced change in Fermi surface topology when the magnetic field suppresses the Kondo-screening induced itineracy of $\rm f$ electrons and is generally indicative of the proximity to a meta-magnetic first order quantum critical point (QCP)~\cite{Kim1}. It should be noted that there is a small positive MR very near $\rm T^{\rm a}$ (Fig.~\ref{fig3}b). However, $\rm T^{\rm a}$ is not necessarily associated with the coherence temperature. Participation of $\rm f$ electrons in the construction of Fermi surface at temperatures much higher than the coherence temperature is not surprising, particularly in the light of recent ARPES studies on the Kondo lattice system $\rm CeCoIn_5$~\cite{Jang}.

\begin{figure}
\includegraphics[width=\linewidth]{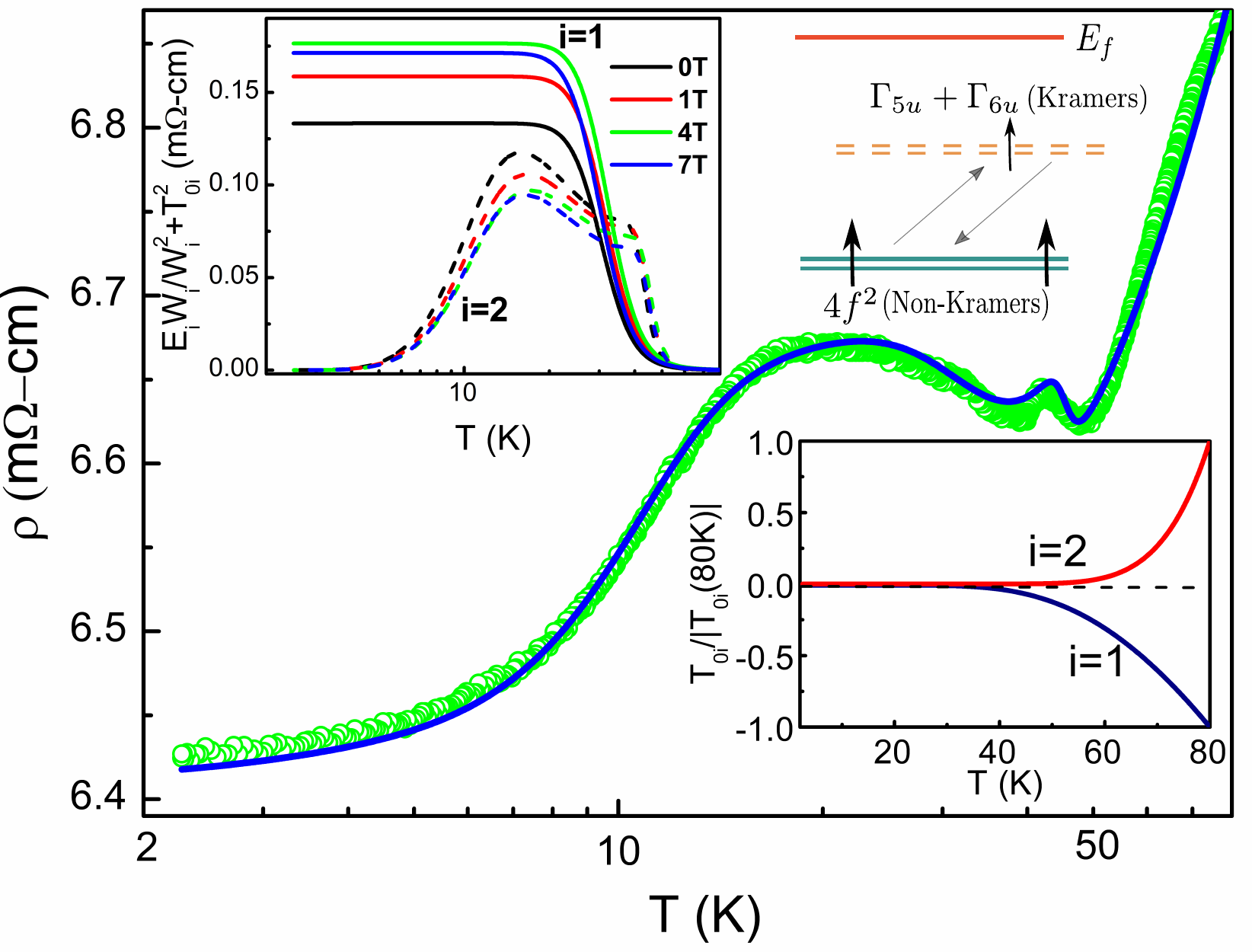}
\caption{The temperature dependence of resistivity of the nanorod fitted using Eq.~\ref{eqn1}. Upper insets:(left) Evolution of the resistivity corresponding to the two hybridization channels (in Eq.~\ref{eqn1}) with the magnetic field. The de-convoluted resistivity suggests existence of the two fluid scenario in PIO nanorod. Ordinary Kondo effect (shown by solid line) occurs when conduction electrons couple to the pseudospin 1/2 degrees freedom corresponding to $\rm 4f^{2}$ non-Kramers doublet. One can also envisage charge transfer between Pr and Ir via the intermediate Kramer's doublet state (shown in the schematic (right)). Excitation of an electron to the Kramer's doublet necessarily leads to the development of a coherent hybridization that breaks both single and double time reversal symmetry, resulting a hastatic order phase (represented by dashed line). Lower inset: The temperature dependence of $\rm T_{\rm 0i}$'s. Both $\rm T_{\rm 01}$ and $\rm T_{\rm 02}$ remains effectively constant at small negative values below $\rm T^\ast$. Interestingly, $\rm T_{\rm 02}$ undergoes sign inversion from negative to positive value at $T^a$.}\label{fig5}
\end{figure}

What is the origin of the resistive anomaly and the low temperature macroscopic Fermi liquid state? Continuing from our discussion earlier, the crystal electric field (CEF) ground state of $\rm Pr^{3+}$ in a pyrochlore lattice is a non-Kramers doublet with a magnetic dipole moment along the local $<111>$ direction and a quadrupole moment in the plane perpendicular to the local $<111>$ direction. For the PIO, the first excited crystal field level is separated from the ground state by a fairly large gap $\sim$160 K and virtual fluctuation to the excited singlet can be neglected. The charge transfer between the Pr and Ir atoms could be mediated through the virtual or physical process (such as oxygen mediated hopping) and necessarily involves intermediate states such as $\rm 4f^{1}$ or $\rm 4f^{3}$. For simplicity, we consider $\rm 4f^{1}$ in our case. In the $\rm D_{3d}$ CEF, an intermediate Kramers pair $\Gamma_{5u}$ and $\Gamma_{6u}$ are developed for the $\rm 4f^{1}$ configuration~\cite{Rau}. The perfectly degenerate two channel Kondo effect mixes the half-integer spin of the intermediate Kramer's doublet with the integer spin state of the the $\rm  4f^{2}$ non-Kramer's doublet. The hybridizations for the two screening channels can be represented by a two component spinor. At the onset of Kondo coherence, the spinorial hybridization breaks the macroscopic spin rotation symmetry as well as single and double time reversal symmetry, leading to a phase transition to the so called `hastatic order'~\cite{Flint,Rau}. In addition, the conduction electrons that directly hybridize with $\rm Pr^{+3}$ ion can give an effective spin 1/2 Hamiltonian which eventually leads to an ordinary Kondo effect similar to reported earlier~\cite{Nakatsuji}. Considering these two distinct hybridizations, one can construct a two-fluid model. The corresponding model Hamiltonian has been discussed in the Appendix B.
%the Hamiltonian can be written as: $\mathcal{H}_{hyb} = \mathcal{H}_{1} + \mathcal{H}_{2} + \text{H.c.}$ where,
%$\mathcal{H}_{1} = \mathcal{J}^1 \sum_{j} c^{\dagger}_{j\alpha}  \vec{\sigma}_{\alpha \beta} c_{j \beta}.\vec{S}_{j}$ and $\mathcal{H}_{2} = $\mathcal{J}^2 \sum_{r,r'} c^{\dagger}_{r \bar{\alpha}} | \Gamma_{5u} \sigma \rangle_{r'} \langle \alpha |_{r'} + \text{Time reversed partner}$.

The correlation between resistivity of Kondo lattice systems and the quasi-elastic linewidth in neutron scattering, usually described by a Lorentzian for $\rm 4f$ bands, is well established~\cite{Freimuth, Garde, Bikash2}. It should be noted that the $\Gamma_3$ doublet ground state has quadrupolar degrees of freedom, too. For simplicity, we neglect the contributions of quadrupolar Kondo effect~\cite{Cox, Kim2} and use two separate Lorentzians to represent the two distinct hybridization channels. The temperature dependence of resistivity could then be written as follows:
\begin{equation}
\rm  \rho(T)= \rho_{0} + aT + E_{1}\frac{W_{1}}{W_{1}^{2}+ T_{01}^{2}}+ E_{2}\frac{W_{2}}{W_{2}^{2}+ T_{02}^{2}} \label{eqn1}
\end{equation}
The parameters $\rm  W_{\rm i}=T_{\rm fi}\exp(-T_{\rm fi}/T)$ and $\rm T_{0i}=A_{\rm i} + B_{\rm i}\exp(-T_{\rm mi}/T)$ are, in general, temperature dependent while $\rm a$, $\rm E_{\rm i}$, $\rm A_{\rm i}$, $\rm B_{\rm i}$, $\rm T_{\rm fi}$ and $\rm T_{\rm mi}$ ($i=1,2$) are constant parameters. For details of the analysis, see Appendix B. The same phenomenological model with single Lorentzian cannot reproduce the observed temperature dependence of resistivity. However, once we add an extra hybridization channel, we find that the resistivity is accompanied by two minima followed by a quadratic drop at low temperature, as shown in Fig.~\ref{fig5}.

The hybridization represented by the quasielastic linewidth $\rm T_{\rm f1}$  follows logarithmic behavior as predicted by Hamann (upper inset, Fig.~\ref{fig5}) and it is certainly conceivable that it is coming from the ordinary Kondo coupling discussed earlier. The Kondo temperature $\rm T_{K}$ estimated from this component turns out to be $29$ K, close to bulk single crystal~\cite{Nakatsuji}. The low temperature T$^{2}$ behavior and the resistivity anomaly at $\rm T_{\rm 0}$ are connected with the formation of an itinerant heavy-electron fluid, as a consequence of mixing between half-integer spin of Kramer's doublet with an integer-spin non-Kramer's doublet in 4f$^{2}$. The resistivity anomaly arises when T$_{02}$ changes sign (see Appendix B). At low temperatures, it thus appears that the resistivity of the PIO nanorod is consistent with a two-fluid description: one is the single ion Kondo impurity fluid, and other is the coherent heavy fermion fluid, characterized by hastatic order. The positive contribution of ordinary Kondo effect (channel $i=1$) and negative contribution of hastatic order (channel $i=2$) in magnetoresistance (inset,~\ref{fig5}) are consistent with the reported data~\cite{Nakatsuji,Kim1}.

%The appearance of the resistive anomaly (or the double minima) and its position depend on the competing nature of the two hybridizations. We construct a diagram describing different possible regimes of electrical conduction within the experimental temperature range allowed by Eq.~\ref{eqn1} (upper inset, Fig.~\ref{fig5}). The three distinguishable scenarios of the temperature dependence of resistivity depending upon the relative contribution of $\rm B$'s and $T_{\rm m}$'s of the two Lorentzians, under the constraint that other parameters used in Eq.~\ref{eqn1} remain fixed at their original best fit value, are as follows (lower inset, Fig.~\ref{fig5}): $1)$ Regime I exhibits a single low temperature minimum; $2)$ Regime II reproduces the resistive anomaly (the double minima); $3)$ Regime III shows single minimum in resistivity at higher temperature compared to Regime I. All other essential features, such as the low temperature Fermi liquid like state and the high temperature metallic behavior remain unchanged in these three regimes. For a given value of $T_{\rm m2}/T_{\rm m1}$, when ${\rm B}_{2}/{\rm B}_{1}$ is small (i.e, channel $1$ dominates over channel $2$), there appears a single minimum at lower temperature (Region I). In region II, we observe two minima in resistivity. The other limit is ${\rm B}_{2}>>{\rm B}_{1}$. This limit corresponds to the situation where channel $2$ dominates over channel $1$ (Region III). More information on the influence of various parameters has been given in Ref.~\cite{SM}.

What is the origin of negative high field MR at low temperature? One possibility is that it could arise due to the disorder in a Kondo lattice, either because of randomly defined $\rm T_K$ at each Pr-site or non-magnetic impurities replacing Pr-sites, giving rise to negative MR at high field~\cite{Ohkawa}. However, as shown in Fig.~\ref{fig3}b, the high field negative MR disappears above $\rm 10 K$, much below the Pr-Pr interaction energy scale $20$ K~\cite{Nakatsuji}, to reappear again around $\rm T^{\rm a}$ and above $\rm T_K$. The negative MR takes maximum value above $\rm T_K$. Such a behaviour cannot be explained within the disordered Kondo lattice framework. Moreover, the quadratic field dependence of negative MR is characteristic of itinerant spin fluctuation~\cite{Moriya, Mukhopadhyay} which is consistent with `hastatic' order having Ising like quasiparticle excitation. Moreover, the nanorod has a large aspect ratio of $10$ and hence the shape anisotropy cannot be neglected. The global Ising axis due to the shape anisotropy is along the symmetry [111] axis of the nanorod, thus further stabilizing the Ising quasiparticles. The suppression of spin fluctuations of the Ising quasiparticles under a magnetic field possibly leads to $\rm H^2$ dependence of negative MR.

Could the coherent Fermi liquid state characterized by quadratic temperature dependence be an artefact of disorder or non-stoichiometry? Indeed, thermodynamic and transport measurements in stuffed (Pr-rich) $\rm Pr_{2+x} Ir_{2-x} O_{7-\delta}$~\cite{Kimura1, Kimura2} reveal a phase transition below $1$ K at ambient pressure, where the resistivity is actually reduced with further lowering of temperature, supposedly due to spin ice order of local $\rm Pr^{3+}$ moments, a behavior which is not observed in stoichiometric bulk sample. However, in the present case, the single crystal nanorod is not Pr rich. Rather it is marginally Pr deficient as suggested by averaging over several EDS line scans. Additionally, the relevant temperature scale in our case is too high to be caused by $\rm Pr$ stuffing.

%What is the origin of enhanced Kondo coupling strength observed in the single crystal nanorod compared to the bulk single crystal? Could it be related to the compressive strain along [111] due to reduced lattice constant? It has already been observed that a tensile strain along [111] direction leads to magnetic long range ordering~\cite{Ohtsuki}. In the present case, the compressive strain along [111] might be leading to enhanced Kondo coupling. In the parameter space of Kondo coupling strength $K$ and and frustration parameter $Q$, a zero temperature quantum phase transition from a metallic spin liquid with small Fermi surface (under-screened local moments) to a heavy fermi liquid with large Fermi surface is possible despite high $Q$, if $K$ is also high at the same time~\cite{Coleman}.

We summarize the findings in the following. The electrical transport properties of $\rm Pr_2 Ir_2 O_7$ single crystal nanorod shows existence of a coherent Fermi liquid state at low temperature with its characteristic $\rm T^2$ behavior of resistivity. 
Such experimental observation of Fermi liquid phase for a Kondo system with non-Kramers doublet ground state itself suggests that it must be of symmetry breaking type, which is further supported by our magnetoresistance measurement. Such non-trivial temperature dependence of resistivity in the present system can be explained incorporating two distinct hybridization channels and such two-fluid description is also consistent with the possibility of existence of a `hastatic' Fermi liquid state in the nanorod.

SM acknowledges Department of Science and Technology (DST), India, for financial support.

\renewcommand{\thefigure}{A\arabic{figure}}
\setcounter{figure}{0}
\renewcommand{\theequation}{A\arabic{equation}}
\setcounter{equation}{0}
\section*{Appendix} 
\begin{figure*}
	\includegraphics[width=0.55\textwidth]{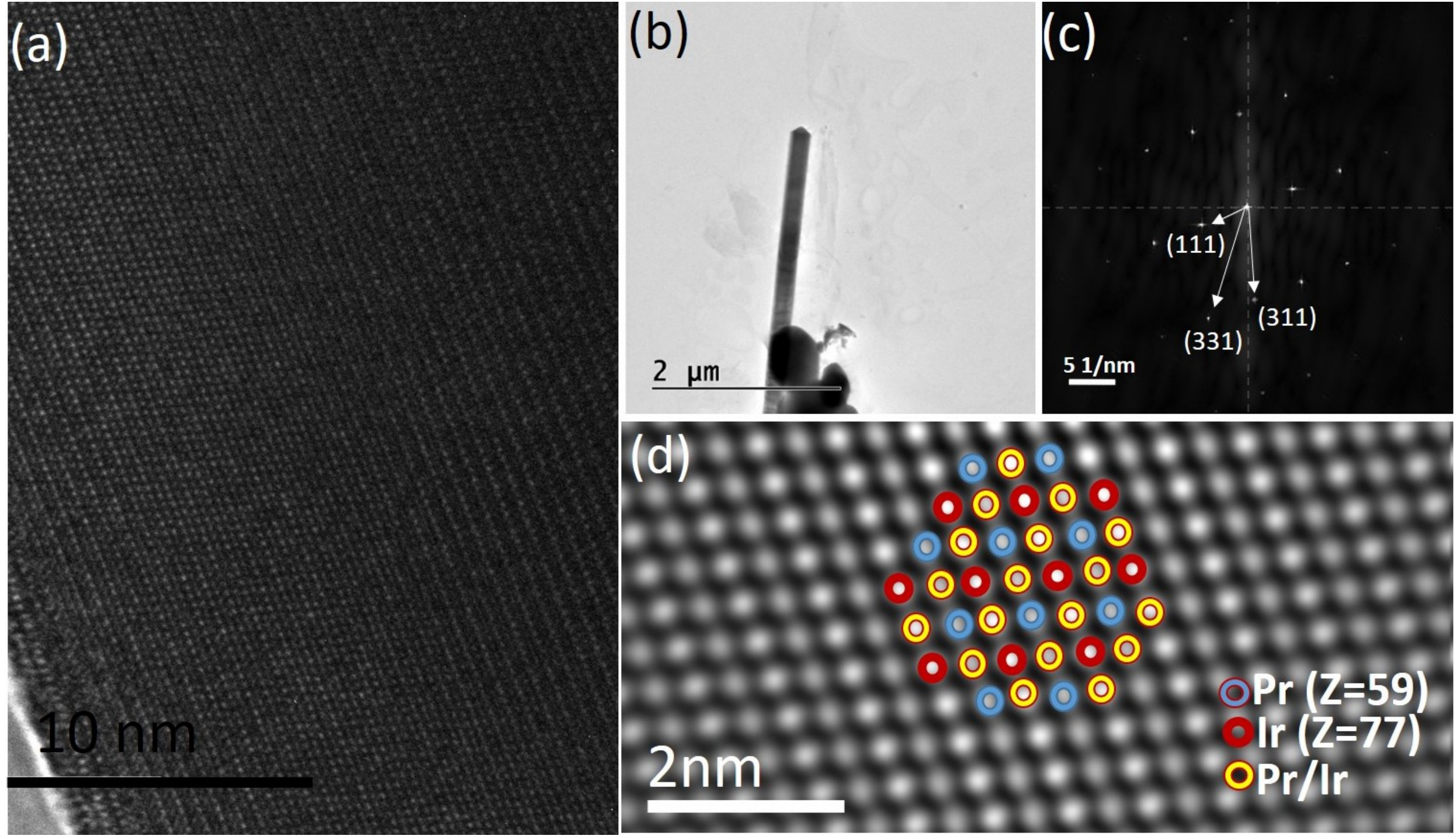}
	\caption{(a) High resolution TEM image of a PIO nanorod showing the crystal growth along $<111>$ . (b) TEM image of a PIO nanorod. (c) SAED pattern of a PIO nanorod showing distinct Bragg spots. (d) Magnified TEM image after Fourier Filtering shows pyrochlore ordering of Pr and Ir sublattices. Pr and Ir atoms are drawn in blue and red color while yellow is used for alternating Pr/Ir atom.
	}\label{BG_Sfig1}
\end{figure*}
\section*{A. EXPERIMENTAL DETAILS: SAMPLE PREPARATION, CHARACTERIZATION AND DEVICE FABRICATION} \label{App-A}
The polycrystalline samples based on $ \rm Pr_2 Ir_2 O_7 $ were prepared following standard solid state route. $ \rm Pr_{6} O_{11} $ (Alfa-Aesar, 99.99$\%$) and $\rm IrO_{2}$ (Alfa-Aesar, 99.99$\%$) powders were mixed in their stoichiometric ratios and then thoroughly ground for several hours. The resulting mixtures were pelletized and heated in air at 1050$^\circ$C for one week with several intermediate grinding. Before every intermediate grinding an additional 5$\%$ $\rm IrO_{2}$ was added in order to compensate the loss during heating. We find that a $5\%$ surplus of $\rm IrO_{2}$ is the optimum amount to obtain stoichiometric $ \rm Pr_2 Ir_2 O_7 $ (PIO). After the final annealing, the sample was initially cooled at 3~K/h down to 800$^\circ$C, and subsequently at 20~K/h down to room temperature. Large number of loosely bound tiny crystals were formed with varying length and cross sectional area on the surface of the pellet, due to the self flux growth involving IrO$_{2}$~\cite{Manni}. The dimension of the crystals were measured using SEM (scanning electron microscope) and optical profilometer. The cross-section of the single crystal nanorods varied between 0.01-0.04 $\mu \rm m^{2}$. We found larger sized micro-crystals too with cross-section varying between 50-1000 $\mu \rm m^{2}$. The typical length of the crystals varied between 10-100 $\mu$m.

A diluted solution containing PIO nanorods in isopropanol was prepared and subsequently drop-casted on marked silicon oxide substrate. The four-probe contact pads were drawn on a suitable crystal using e-beam lithography and subsequent metallization with Au/Cr using e-beam evaporator. On the other hand, larger sized micro-crystals were transferred directly on silicon oxide substrate and contacted with silver epoxy for transport measurement.

Transmission electron microscopy (TEM) of the PIO nanorods were performed at room temperature using a FEI Titan G2 60 -300 for structural investigation. The TEM samples were prepared by drop casting the solution containing PIO nanorods on a copper grid coated with carbon film. Fig.~\ref{BG_Sfig1}(a) displays high resolution TEM image of a PIO nanorod shown in Fig.~\ref{BG_Sfig1}(b). The periodic contrast between two column of atoms appear due to Pr and Ir atomic planes. Close inspection of TEM and selected area electron diffraction (SAED) images reveal the single crystalline nature of PIO nanorod which preferentially grows along the easy axis $<111>$. Fig.~\ref{BG_Sfig1}(d)shows Z-contrast of the individual atoms representing Pr and Ir. The most intense spots are responsible for $^{77}\rm Ir$  and less intense one for $^{59} \rm Pr$ because in HRTEM image, the contrast is proportional to the atomic number. Thus the HRTEM image gives clear evidence of pyrochlore order.

\subsection*{X-ray photoelectron spectroscopy (XPS) measurement}
\begin{figure*}
	\includegraphics[width=0.8\linewidth]{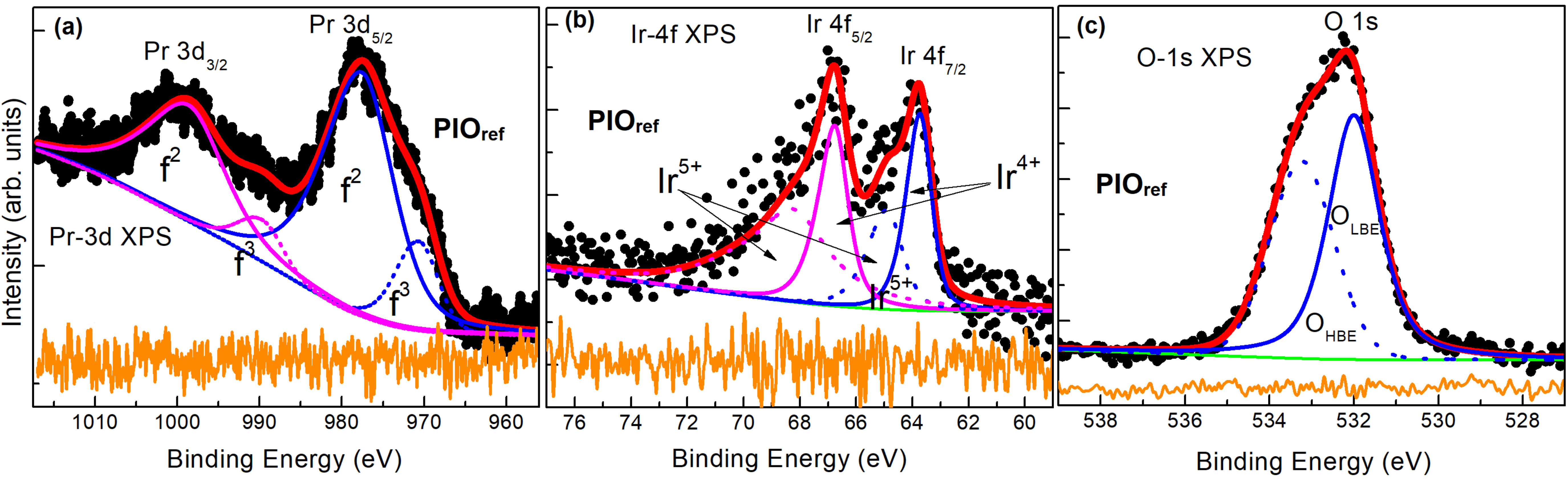}
	\caption{The XPS spectra of (a)  Pr 3d  (b) Ir 4f and (c) O 1s for a PIO$_{\rm ref}$ sample.
	}\label{BG_Sfig2}
\end{figure*}

Fig.~\ref{BG_Sfig2}(a)-(c) shows de-convoluted XPS spectra of a representative micro-crystal PIO$_{\rm ref}$. Interestingly, the whole Pr 3d XPS spectra [Fig.~\ref{BG_Sfig2}(a)] of single crystal PIO are shifted towards higher binding energies $\sim$ 45~eV along with broadening in peak widths as compared to its bulk poly-crystalline PIO Pr 3d XPS~\cite{Bikash2}. We have repeated the XPS measurement several times on the same crystal and different crystals too, but the results were the same. Qualitatively, the spectral character for Pr 3d is almost similar in single crystal and bulk poly-crystalline PIO~\cite{Bikash2}.

In single crystalline PIO [Fig.~\ref{BG_Sfig2}(a)], the two main XPS peaks termed as $ \rm Pr 3d_{5/2}$ and $ \rm Pr 3d_{3/2}$ states are centered at binding energies $\sim$977~eV and $\sim$998~eV. The satellite peaks are situated at $\sim$970~eV and$\sim$ 990~eV. Here the two peaks centered at binding energies $\sim$977~eV (main peak) and $\sim$970~eV (satellite peak) are labelled as $\rm f^2$ and $\rm f^3$ states, respectively~\cite{Gamza}. The ratio of $\rm f^3$/$\rm f^2$ is found to be  0.21 for the $ \rm Pr 3d_{5/2}$ state. This suggests existence of mixed oxidation states of Pr-ion at room temperature. We have also calculated the value of hybridization strength~\cite{Gamza} between the Pr 4f orbital and the itinerant Ir 5d conduction electrons, which turns out to be  0.17 eV , suggesting the presence of strong hybridization between the Pr 4f orbital and conduction electrons. The mixed oxidation states of Pr are possibly due to valence fluctuation. These are characteristics of Kondo-like systems~\cite{Gamza,Neumann}. Fig.~\ref{BG_Sfig2}(b) shows Ir 4f core-level spectra with asymmetric lines. Two main peaks Ir $\rm 4f_{7/2}$ and Ir $\rm 4f_{5/2}$ are observed due to spin-orbit splitting$\sim$ 3~eV. The de-convoluted Ir 4f XPS spectra suggest the mixed oxidation states of Ir. Fig.~\ref{BG_Sfig2}(c) shows O 1s spectra consisting of two peaks. The lower energy peak arises due to O$^{-2}$ anion in the system and higher energy peak is possibly associated with the presence of $\rm O$-Pr$^{+4}$~\cite{Paunovic}.

%% \textcolor{blue}{\textit{}}

\section*{B. Analysis of resistivity data} \label{App-B}

In this section we discuss possible phenomenological origin of the symmetry broken Fermi liquid phase at low temperature, where the resistivity data shows $T^2$ behavior at low temperature along with possible magnetic ordering. We analyze below that \textit{two} temperature dependent Lorentzians are required to fit the data accurately. Explanation of such  behavior is proposed to be originating from two distinct hybridization processes and the aforementioned phase can be attained by means of a symmetry breaking hybridization similar to the theory proposed by Chandra \textit{et al}.~\cite{chandra}.

\begin{figure*}
	\includegraphics[width=\linewidth]{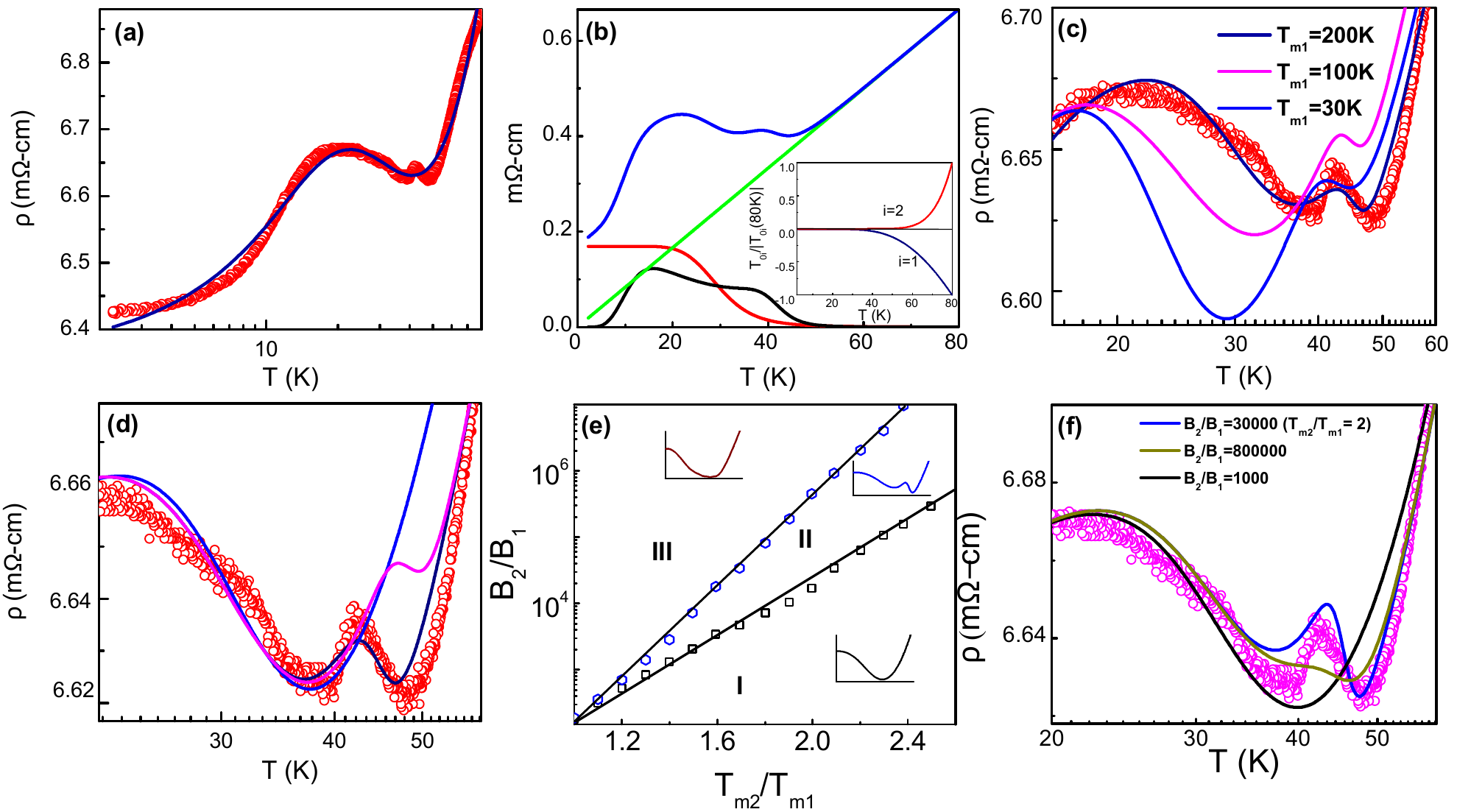}
	\caption{(a) The temperature dependence of resistivity of PIO nanorod fitted using Eq.~\ref{eqn2}. The Eq.~\ref{eqn2} with single Lorentzian cannot reproduce the resistive anomaly as well as low temperature $\rm T^{2}$ behavior. (b) The temperature dependence of individual contribution to the resistivity (blue) of the three terms in Eq.~\ref{eqn3}. The variation of aT, $\rm E_{1}\frac{W_{1}}{W_{1}^{2}+ T_{01}^{2}}$ and $\rm E_{2}\frac{W_{2}}{W_{2}^{2}+ T_{02}^{2}}$ are shown in green, red and black, respectively. Inset: The reduced temperature derivative of $\rm T_{\rm 0i}$'s. Both $\rm T_{\rm 01}$ and $\rm T_{\rm 02}$ remains constant roughly where $\rm T^{\rm2}$ dependence is followed. (c) The effect of $\rm T_{\rm m1}$ (in Eq.~\ref{eqn3}) on the resistivity, showing lower temperature minimum shifts towards lower $\rm T$ and higher temperature minimum gradually disappears with decreasing value of $\rm T_{\rm m1}$ from its best fit value ($\rm T_{\rm m1}$=200~K). (d) The effect of $\rm T_{\rm m2}$ (in Eq.~\ref{eqn3}) on the resistivity, showing that with further increase of $\rm T_{\rm m2}$ from its best fit value ($\rm T_{\rm m2}$=370~K), the higher temperature minimum gradually disappears but the lower temperature minimum remains unchanged. Beyond a threshold value of $\rm T_{\rm m2}$ only lower temperature minimum survives and the resistivity becomes insensitive to  $\rm T_{\rm m2}$. (e) Different transport regimes as predicted by Eq.~\ref{eqn3}, depending on the relative contribution of ${\rm B}_{2}/{\rm B}_{1}$ and ${\rm T}_{\rm m2}/{\rm T}_{\rm m1}$, are labeled as I, II and III (discussed in the text). (f) The corresponding simulated resistivity for the three regimes are shown by continuous lines by varying ${\rm B}_{2}/{\rm B}_{1}$ for a fixed value of ${\rm T}_{\rm m2}/{\rm T}_{\rm m1}$.}\label{BG_Sfig3}
\end{figure*}

\subsection*{The two fluid description model}
The two fluid description is an outcome of two separate hybridization mechanisms taking place between the Ir conduction electrons and Pr impurity ions. These two processes are defined by the Hamiltonians $\mathcal{H}_1$ and $\mathcal{H}_2$, that we briefly discuss below. 

The Praseodymium ground state in $ \rm Pr_2 Ir_2 O_7 $ with $\rm Fd\bar{3}m$ symmetry is $\rm 4f^{2}$, having quantum number $ L\rm=4$, $S\rm=1 $ and $ J\rm=4$. This energy level is nine fold degenerate and the cubic crystal electric field effects (CEF) lift the degeneracy resulting in three doublets and three singlets. The ground state of $\rm 4f^{2}$ multiplet is a $\Gamma_{3}$ non-Kramers doublet separated from the first excited state, which is singlet, by a fairly large gap of $\sim$160 K. Since, the CEF level separation is large, the virtual fluctuations from the ground state to excited states is supposedly negligible. The intrinsic doubly degenerate $\rm 4f^{2}$ configuration can be effectively described in terms of pseudospin 1/2 degrees of freedom. Onoda \textit{et al}.~\cite{onoda} theoretically showed that the systems like $ \rm Pr_2 Ir_2 O_7 $ can be expressed by an effective pseudospin-1/2 model. The high temperature bulk susceptibility measurement on $\rm Pr_2 X_2 O_7 $ ($\rm X$=Sn, Hf) reveals that the 4f moments embedded within the conduction sea are Ising in nature ~\cite{Princep,Sibille}. In this case, the quasiparticle hybridization of the impurity pseudospin with the conduction electrons leads to an ordinary Kondo effect where no fundamental symmetry is broken. This is identified as hybridization of first kind for our paper. The Hamiltonian corresponding to this hybridization can be written as follows:
\begin{equation}\label{Mod_Kon_Con}
	\mathcal{H}_{1} =  \mathcal{J}_1 \sum_{j} c^{\dagger}_{j\alpha}  \vec{\sigma}_{\alpha \beta} c_{j \beta}.\vec{S}_{j}
\end{equation}
Where $\mathcal{J}_1$ is the coupling constant and $c^{\dagger}_{j\alpha}$ represents creation operator for conduction electron with spin $\alpha$. Spin 1/2 Pauli matrices $\vec{S}_{j}$ and $\vec{\sigma}_{j}$ denote impurity pseudospin and conduction electron spin, respectively. The $\mathcal{H}_{1}$ which preserves time reversal symmetry contributes to the logarithmic behavior in the resistivity.

\begin{table*}
	\centering
	\caption{Fitting parameters in the phenomenological model (Eq.~\ref{eqn3}) describing temperature dependence of resistivity at different magnetic field values discussed in the main text. Recall that the application of magnetic field changes the positions of the resistive anomaly without changing the low temperature ${\rm T}^2$ behavior as well as the high temperature resistivity.}\label{T1}
	\begin{tabular}{c c c c c c c c c c c c c c c c c c}
		\hline 
		$\rm B$ &$\rho_{0}$ & $\rm a$& $\rm E_{1}$ & $\rm T_{\rm f1}$ & $\rm A_{1}$ & $\rm B_{1}$ & $\rm T_{\rm m1}$ & $\rm E_{2}$ & $\rm T_{\rm f2}$ & $\rm A_{2}$ & $\rm B_{2}$ & $\rm T_{\rm m2}$ \\
		(T)  & (m$\Omega$ cm) & (m$\Omega$ cm/K) & (m$\Omega$ cm K) & (K)  & (K)  &(K)   & (K) & (m$\Omega$ cm K) & (K) & (K) & (K) & (K) \\
		\hline
		0 & ~6.2 & ~0.008 & ~0.14 & ~3$\times$10$^{-4}$ & ~-0.018 & -6.9 & 210 & 1.07 & 29.68 & ~-4.55 & 8.4$\times 10^{4}$ & 370 \\
		
		1 & ~6.2 & ~0.008 & ~0.14 & ~5$\times$10$^{-4}$ & ~-0.021 ~&  ~-7.1  & 213 & 1.06 & ~29.53 & ~-5.01 & 8.4$\times 10^{4}$ &  379 \\
		
		4 & ~6.2 & ~0.008 & ~0.13 & ~4.4$\times$10$^{-4}$ &~ -0.018 & ~-6.4 & ~219 & 1.07 & ~29.90 & ~-5.21 & 7.80$\times 10^{4}$  & 383 \\
		
		7 & ~6.2 & ~0.008 & ~0.14 & ~5.4$\times$10$^{-4}$ & ~-0.021 & ~-6.34 & ~200 & 0.99  & ~30.11 &-4.81 & 9.90$\times 10^{4}$ & 390 \\
		
		\hline
	\end{tabular}
\end{table*}

One note that $\mathcal{H}_{1}$ does not exhaust all the possiblities of  hybridization. Rau \textit{et al}.~\cite{Rau} theorized possibility of another kind of hybridization in $\rm Pr_2 Ir_2 O_7$, which is similar to the developement of hastatic order proposed by Chandra \textit{et al}.~\cite{chandra}. In contrast to $\mathcal{H}_1$, here the charge transfer between Pr and Ir takes place via an intermediate Kramer's doublet state, originating from hopping between conduction electrons of Ir5d $j_{\rm eff}$=1/2 state and Pr4f moment. This process can occur via several mechanisms such as oxygen mediated hoppings. Such hybridization has a significant implication on the Kondo system with non-Kramers ground state. The non-Kramer's $\Gamma_{3}$ doublet of $\rm Pr^{3+}$ in $ \rm Pr_2 Ir_2 O_7 $ has a magnetic dipole moment along the $<111>$ direction and quadrupolar moment in the plane perpendicular to the local $<111>$ direction. The charge transfer between the Pr and Ir atom through the physical or virtual processes necessarily involves intermediate states such as $\rm 4f^{1}$ or $\rm 4f^{3}$, both of which has Kramer's configurations. The $\rm 4f^{1}$ state is more likely, being lower in energy. In the $\rm D_{3d}$ CEF, a Kramer's pair $\Gamma_{5u}$+$\Gamma_{6u}$ is developed for the $\rm 4f^{1}$ configuration, given by $ m = \pm 3/2$ states of $j=5/2$ manifold~\cite{Rau}. The quasiparticle mixing of half-integer-spin in Kramer's doublet with an integer-spin in the non-Kramer's doublet results in a two channel Kondo effect. We follow the same Hamiltonian modeled by Rau \textit{et al}.~\cite{Rau} :
\begin{equation}\label{Mod_Kon_Symm}
	\mathcal{H}_{2} =  \sum_{r,r'} \mathcal{J}_2(r,r',\beta) c^{\dagger}_{r \beta} | \Gamma_{5u} \rangle_{r'} \langle \Gamma_{3} \bar{\beta} |_{r'}+\text{time reversed} +\text{H.c.}
\end{equation}
Where $r$ is an Ir site and $r'$ is a Pr site. We note $\mathcal{J}_2(r,r',\beta)$ is a constant that depends on the microscopic details of the system and $\beta$ labels the integer spin state of non-Kramers doublet, with $\rm \bar{\beta}=-\beta$.  In the present case, the hybridization between conduction electron and impurity pseudospin results in mixing between spin half and spin integer states. Half-integer and integer spins transform differently under time reversal symmetry operation. Mean field calculation \cite{MFhyb} shows that the hybridization of type $\mathcal{H}_2$, breaks both the single and the double time reversal symmetry, similar to the hastatic order proposed for $\rm URu_{2} Si_{2}$~\cite{chandra}.

In the following we discuss how these two distinct hybridization processes are related to resistivity. In Eq.~\ref{eqn3} we will show the two Lorentzian functions required to fit the data, as mentioned earlier. Essentially these two functions imply two types of Kondo like contributions in the resistivity. One Lorentzian arises from hybridization process of the form $\mathcal{H}_1$. The resistivity corresponding to this type of term has been previously obtained by Hamann \cite{Hamann}. The second Lorentzian shows departure from the behavior predicted by Hamman. Its non-trivial dependence on temperature results in $\rm T^2$ functional form of the resistivity at low temperature. We argue that the underlying physics of such non-trivial dependence is connected to symmetry breaking due to mixing between conduction electron and impurity spin, governed by Hamiltonian of the form $\mathcal{H}_2$ as shown in Eq.~\ref{Mod_Kon_Symm}.

\subsection*{Lorentzian fitting}

In a Kondo system, the dominant contribution to the resistivity at low temperature arises from the scattering between conduction electrons and narrow Lorentzian shaped 4f band. The 4f electron in $\rm Pr^{+3}$ ion is well localized and the effective 4f density of states at the Fermi level could be expressed as~\cite{Freimuth,Garde},
\begin{equation}
	\rm N(E_{f})= \frac{W}{W^{2}+(E_{\rm F}-E_{\rm f})^{2}} \label{eqn1}
\end{equation}
The magnetic contribution to the resistivity is proportional to the $\rm N(E_{\rm f})$. Here, $\rm W=T_{\rm f} exp(-T_{\rm f}/T)$, $\rm T_{\rm f}$ is a constant parameter which is identical with the quasi-elastic line-width of the neutron spectra. Only those 4f states which are in proximity to the Fermi level, take part in the scattering process and contribute in the resistivity , at low temperature. The position of the center of gravity of the 4f level with respect to Fermi level can be expressed as $\rm K_{\rm B}T_{r}=(E_{ F}-E_{\rm f})$. In this model $\rm T_{r}$ is, in general, temperature dependent and is given by, $\rm T_{r}= A + B\exp(-T_{\rm m}/T)$, where A, B and $\rm T_{\rm m}$ are the constants for a given compound. $\rm T_{\rm m}$ is related to the crystal field excitation~\cite{Freimuth,Garde}. The electrical resistivity can be written as,
\begin{equation}
	\rho(T)\rm = \rho_{0} + aT + E\frac{W}{W^{2}+ T_{r}^{2}} \label{eqn2}
\end{equation}

Here, $\rho_{0}$ is the residual resistivity and the linear term in $\rm T$ represents the phononic contribution to the resistivity. However, in the present case, the resistivity given by Eq.~\ref{eqn2} cannot reproduce the resistive anomaly as well as low temperature $\rm T^{2}$ behavior as shown in  Fig.~\ref{BG_Sfig3}a. To reproduce the nanorod resistivity we must include an additional Lorentzian and the total resistivity can be written as follows:

\begin{equation}
	\rho(T)= \rm  \rho_{0} + aT + E_{1}\frac{W_{1}}{W_{1}^{2}+ T_{01}^{2}}+ E_{2}\frac{W_{2}}{W_{2}^{2}+ T_{02}^{2}} \label{eqn3} 
\end{equation}
Where,
\begin{equation} 
	\rm T_{0i}=A_{\rm i} + B_{\rm i}\exp(-T_{\rm mi}/T) \label{eqn4}
\end{equation}
and
\begin{equation}
	\rm W_{\rm i}=T_{\rm fi}\exp(-T_{\rm fi}/T); i=1, 2 \label{eqn5}
\end{equation}

Physically, these two Lorentzians correspond to two different screening channels. The ratio $\rm E_{1}/E_{2}$ represents the relative contributions of the two channels to the magnetic resistivity. We find that the resistive anomaly as well as the low temperature $\rm T^{2}$ regime are described well with Eq.~\ref{eqn3} as shown in the main text. The magnetic field dependent fitting parameters of $\rm \rho(T)$ are tabulated in table~\ref{T1}.

Fig.~\ref{BG_Sfig3}(b) shows temperature dependence of individual contributions to the resistivity of the three terms in Eq.~\ref{eqn3} and the resultant resistivity. The resistivity corresponding to the hybridization represented by quasi-elastic linewidth $\rm T_{\rm f1}$ follows logarithmic behavior as described by Hamann's expression \cite{Hamann}. It is certainly conceivable that this comes from the coupling of the conduction electrons (at the Fermi level) to the 4f moment and gives ordinary Kondo effect where no fundamental symmetry is broken, similar to reported by Nakatsuji \textit{et al}.~\cite{Nakatsuji}. While the low temperature $\rm T^{\rm 2}$ behavior and the resistivity anomaly at $\rm T^{a}$ are contributed by the hybridization corresponding to the quasi-elastic linewidth $\rm T_{\rm f2}$. At $\rm T^{\rm a}$, $\rm T_{\rm 02}$ changes sign from negative to positive [Inset, Fig.~\ref{BG_Sfig3}(b)]. We argue that this hybridization represents the mixing of integer spin state at $\Gamma_{3}$ non-Kramer's doublet to the half-integer spin at $\Gamma_{5u}+\Gamma_{6u}$ Kramer's doublet, as discussed in the main text. This hybridization breaks both single and double time-reversal symmetry, similar to the `hastatic' order.

We find that the positions of the minima are dependent on the parameters $\rm B_{1}$, $\rm B_{2}$ , $\rm T_{\rm m1}$ and $\rm T_{\rm m2}$ only, without changing the high temperature metallic and low temperature $\rm T^{2}$ behavior (see table~\ref{T1}). The parameters $\rm a$, $\rm E_{1}$, $\rm E_{2}$ , $\rm T_{\rm f1}$ and $\rm T_{\rm f2}$ are mainly responsible for altering the value of $\rm \rho(T)$ for a particular temperature $\rm T$. Fig.~\ref{BG_Sfig3}(c) shows the effect of $\rm T_{\rm m1}$ on the resistivity. With decreasing value of $\rm T_{\rm m1}$ from its best fit value, the low temperature minimum shifts towards lower $\rm T$. The $\rm \log T$ dependence in the intermediate temperature range and the high temperature minimum completely disappear below a threshold value. Conversely, for increasing value of $\rm T_{\rm m1}$ from its best fit value, the low temperature minimum eventually disappears and the high temperature one survives. On the other hand, for the increasing value of $\rm T_{\rm m2}$ from its best fit value, the high temperature minimum gradually disappears and beyond a threshold value, only the low temperature minimum remains [Fig.~\ref{BG_Sfig3}(d)]. Moreover, beyond this limiting value, the resistivity becomes insensitive to $\rm T_{\rm m2}$. 

The appearance of the resistive anomaly at $\rm T^a$ depends on the competing nature of the two hybridizations. We construct a diagram describing different possible regimes of electrical conduction within the experimental temperature range allowed by Eq.~\ref{eqn3} (Fig.~\ref{BG_Sfig3}(e)). While constructing the transport-regime diagram, the parameters $\rm a$, $\rm A_{1}$, $\rm A_{2}$, $\rm E_{1}$, $\rm E_{2}$ , $\rm T_{\rm f1}$ and $\rm T_{\rm f2}$ were kept constant and $\rm B_{1}$, $\rm B_{2}$, $\rm T_{\rm m1}$ and $\rm T_{\rm m2}$ were varied. The three distinguishable scenarios of the temperature dependence of resistivity depending upon the relative contribution of $\rm B$'s and $\rm T_{\rm m}$'s of the two Lorentzians, under the constraint that other parameters used in Eq.~\ref{eqn3} remain fixed at their original best fit value, are as follows (Fig.~\ref{BG_Sfig3}(f)): $1)$ Regime I exhibits a single low temperature minimum; $2)$ Regime II reproduces the resistive anomaly (or the double minima); $3)$ Regime III shows single minimum in resistivity at higher temperature compared to Regime I. All other essential features, such as the low temperature Fermi liquid like state and the high temperature metallic behavior remain unchanged in these three regimes. For a given value of $\rm T_{\rm m2}/T_{\rm m1}$, when ${\rm B}_{2}/{\rm B}_{1}$ is small, there appears a single minimum at lower temperature (Region I). In region II, we observe two minima in resistivity. The other limit is ${\rm B}_{2}\gg{\rm B}_{1}$ (Regime III). 

%Fig.~\ref{BG_Sfig3}(d) shows the evolution of the two Lorentzians with magnetic field. The Lorentzian corresponding to channel 2 (dashed line) is quite stable under the magnetic field while the Lorentzian for channel 1 shifts towards higher temperature with increasing magnetic field. The temperature derivatives of the $T_{\rm 0i}$'s are plotted in Fig.~\ref{BG_Sfig3}(e). A broad extremum is developed for the derivative of $T_{\rm 01}$ near $T_K$ and then gradually decreases with decreasing temperature. On the other hand, the temperature derivative of $T_{\rm 02}$ remains close to zero in the temperature range below $T_K$, suggesting that $T_{\rm 02}$ is effectively constant in this temperature regime. Interestingly, both $T_{\rm 01}$ and $T_{\rm 02}$ becomes constant in the Fermi liquid regime.

%While constructing the transport-regime diagram discussed in the main text, the parameters $a$, $A_{1}$, $A_{2}$, $E_{1}$, $E_{2}$ , $T_{\rm f1}$ and $T_{\rm f2}$ were kept constant and $B_{1}$, $B_{2}$, $T_{\rm m1}$ and $T_{\rm m2}$ were varied. We construct the diagram below 300~K for different values of the ratio $T_{\rm m1}/T_{\rm m2}$ and find that the regime boundaries are independent of the ratio [Fig.~\ref{BG_Sfig3}(f)]. The broadening of the boundaries appear because of the error in the value of $B_{1}$ and $B_{2}$ arising out of error in the resistivity minima positions. The diagram shown in the main text is obtained by averaging over $B_{2}/B_{1}$ for each value of $T_{\rm m2}/T_{\rm m1}$.

\end{document}